\documentclass[fleqn,twoside]{article}
\usepackage{espcrc2}

\usepackage{graphicx}
\usepackage[figuresright]{rotating}

\newcommand{\cc}{{\langle\phi\rangle}}

\newcommand{\AmS}{{\protect\the\textfont2
  A\kern-.1667em\lower.5ex\hbox{M}\kern-.125emS}}

\title{Chiral and Critical Behavior in Strong Coupling QCD}
\author{Shailesh Chandrasekharan \address
{Department of Physics, Duke University, Durham NC 27708-0305}
        \thanks{This work was done in collaboration with F.-J.~Jiang and
was supported by the grants \#DE-FG-96ER40945 and 
\#DE-FG02-03ER41241 from the Department of Energy.}}

\begin{document}

\begin{abstract}
We use a cluster algorithm to study the critical behavior of strongly 
coupled lattice QCD in the chiral limit. We show that the finite temperature 
chiral phase transition belongs to the $O(2)$ universality class as
expected. When we compute the finite size effects of the chiral 
susceptibility in the low temperature phase close to the transition, 
we find clear evidence for chiral singularities predicted by chiral 
perturbation theory (ChPT). On the other hand it is difficult
to reconcile the quark mass dependence of various quantities near the 
chiral limit with ChPT.
\vspace{1pc}
\end{abstract}

\maketitle

\section{INTRODUCTION}

One of the important challenges in lattice QCD is to compute quantities
that are dominated by the physics of light quarks. Although there has
been substantial effort in extracting such quantities from the lattice
results by matching the data with ChPT, it is unclear if the current range 
of masses used in the calculations are in the range where ChPT is valid
\cite{Ber02}. Recently, efforts have been directed in two directions: 
a) Improving actions that reduce lattice artifacts. b) Improving ChPT 
that take these artifacts into account. However, we think 
that it is also equally important to find improved algorithms to approach 
the chiral limit.

Over the last decade a new class of cluster algorithms have emerged for
solving a variety of lattice field theories \cite{Eve03}. These algorithms
help in beating critical slowing down very efficiently. Recently, it was
shown that the strong coupling limit of lattice gauge theories with 
staggered fermions can be solved using these new algorithms \cite{Cha03}. 
For the first time, this allows us to study the physics of massless quarks 
on large lattices from first principles. 

In this article we present results from our study of the chiral and
critical behavior near the chiral phase 
transition in strongly coupled lattice QCD with staggered fermions.
We use $U(3)$ gauge fields instead of $SU(3)$ in order to avoid technical
complications in the algorithm. Further details of our study
can be found in \cite{Cha03a}.

\section{MODEL AND OBSERVABLES}

The model we study can be specified by the Euclidean action, 
\begin{eqnarray}
\label{fact}
- \sum_{x,\mu} \frac{\eta_{x,\mu}}{2}\Big[\bar\psi_x U_{x,\mu} 
\psi_{x+\hat{\mu}}
- \bar\psi_{x+\hat{\mu}} U^\dagger_{x,\mu} \psi_x\Big]
\nonumber \\
- m \sum_x \bar\psi_x\psi_x,
\end{eqnarray}
where $m$ is the staggered quark mass, $U_{x,\mu}$ are $U(3)$
link variables and $\psi,\bar\psi$ are Grassmann variables representing
the staggered quark fields. We choose the staggered fermion
phase factors $\eta_{x,\mu}$ to have the property that $\eta_{x,\mu}^2 = 1, 
\mu=1,2,3$ (spatial directions) and $\eta_{x,4}^2 = T$ (temporal direction), 
where the real parameter $T$ acts like a temperature. By working on 
asymmetric lattices with $L_t << L$  and allowing $T$ to vary continuously, 
one can study the finite temperature phase transition in strong coupling QCD 
\cite{Boy92}.

The model described by the partition function $Z(T,m)$, constructed 
from the above action in the usual way, is known to have an 
exact $O(2)$ chiral symmetry when $m=0$. This symmetry is broken at 
low temperatures but gets restored at high temperatures due to a finite 
temperature chiral phase transition. In order to study the chiral physics 
near this transition we focus on the chiral condensate,
\begin{equation}
\langle\phi\rangle = \frac{1}{L^3} \frac{1}{Z}\frac{\partial}{\partial m} 
Z(T,m),
\end{equation}
the chiral susceptibility,
\begin{equation}
\chi = \frac{1}{L^3} \frac{1}{Z}\frac{\partial^2}{\partial m^2} Z(T,m),
\end{equation}
and the helicity modulus,
\begin{equation}
Y_m = \frac{1}{L^3}\Bigg\langle 
\Big\{ \sum_{\mu=1}^3\ \ [\sum_x J_{x,\mu}]^2\Big\}
\Bigg\rangle,
\end{equation}
where $J_{x,\mu} = \sigma_x (b_{x,\mu} - N/8)$, with $\sigma_x = 1$ on
even sites and $\sigma_x = -1$ on odd sites. 
When $m=0$ the current $J_{x,\mu}$ is the conserved current associated with 
the  $O(2)$ chiral symmetry. Further, as discussed in \cite{Has90}, it
can be shown that $F^2 = \lim_{L\rightarrow \infty} Y_m(m=0)$, where $F$ is
related to the pion decay constant. We also measure the Goldstone 
pion mass $M_\pi$.

\section{UNIVERSAL PREDICTIONS}

The predictions of ChPT for $O(N)$ models have 
been discussed in \cite{Has90}. In particular the finite size scaling 
formula for $\chi$ at $m=0$ is given by
\begin{equation}
\chi = \frac{1}{N}\Sigma^2 L^3\Big[ 1 + \beta_1 (N-1) \frac{1}{F^2 L} + 
\frac{a}{L^2} +...\Big],
\label{chptchi}
\end{equation}
where $N=2$ in our case, $\beta_1=0.226...$, 
$\Sigma=\lim_{m\rightarrow 0}\lim_{L\rightarrow \infty} \langle\phi\rangle$,
and $a$ is a constant dependent on other low energy constants. The quark 
mass dependence of $\langle\phi\rangle,\chi$ and $Y_m$ are given by
\begin{eqnarray}
\label{ccm}
\cc &=& \Sigma\Big[1 + \alpha_1 \sqrt{m} + \alpha_2 m + ...\Big], \\
\label{ym}
Y_m &=& F^2[1 + \alpha_3 \sqrt{m} + \alpha_4 m + ...\Big], \\
M_\pi^2 &=& \frac{\Sigma m}{F^2} [1 + \alpha_5 \sqrt{m} + \alpha_6 m 
+ ...\Big].
\label{chptmass}
\end{eqnarray}
For $N=2$ one further finds that $\alpha_3 = 0$ \cite{Has90}. One of the 
consequences of chiral symmetry can be described by the relation
\begin{equation}
\frac{\cc m}{M_\pi^2 F_m^2} = 1 + {\cal O}({m}),
\label{gorr}
\end{equation}
which is the Gellmann-Oaks-Renner relation.

If the chiral phase transition is second order then close to the
critical temperature $T_c$ we have
\begin{eqnarray}
\label{sigma}
\Sigma(T) &=& A (T_c - T)^\beta, \  T < T_c , \\
\lim_{L\rightarrow \infty} \langle\phi\rangle &=& B m^{1/\delta},\  T = T_c,
\label{crit}
\end{eqnarray}
The $O(2)$ universality predicts predicts $\beta=0.3485(2)$, 
$\delta=4.780(2)$ \cite{Cam01}. 

\section{RESULTS}

\begin{figure}[htb]
\vspace{9pt}
\includegraphics[width=16pc]{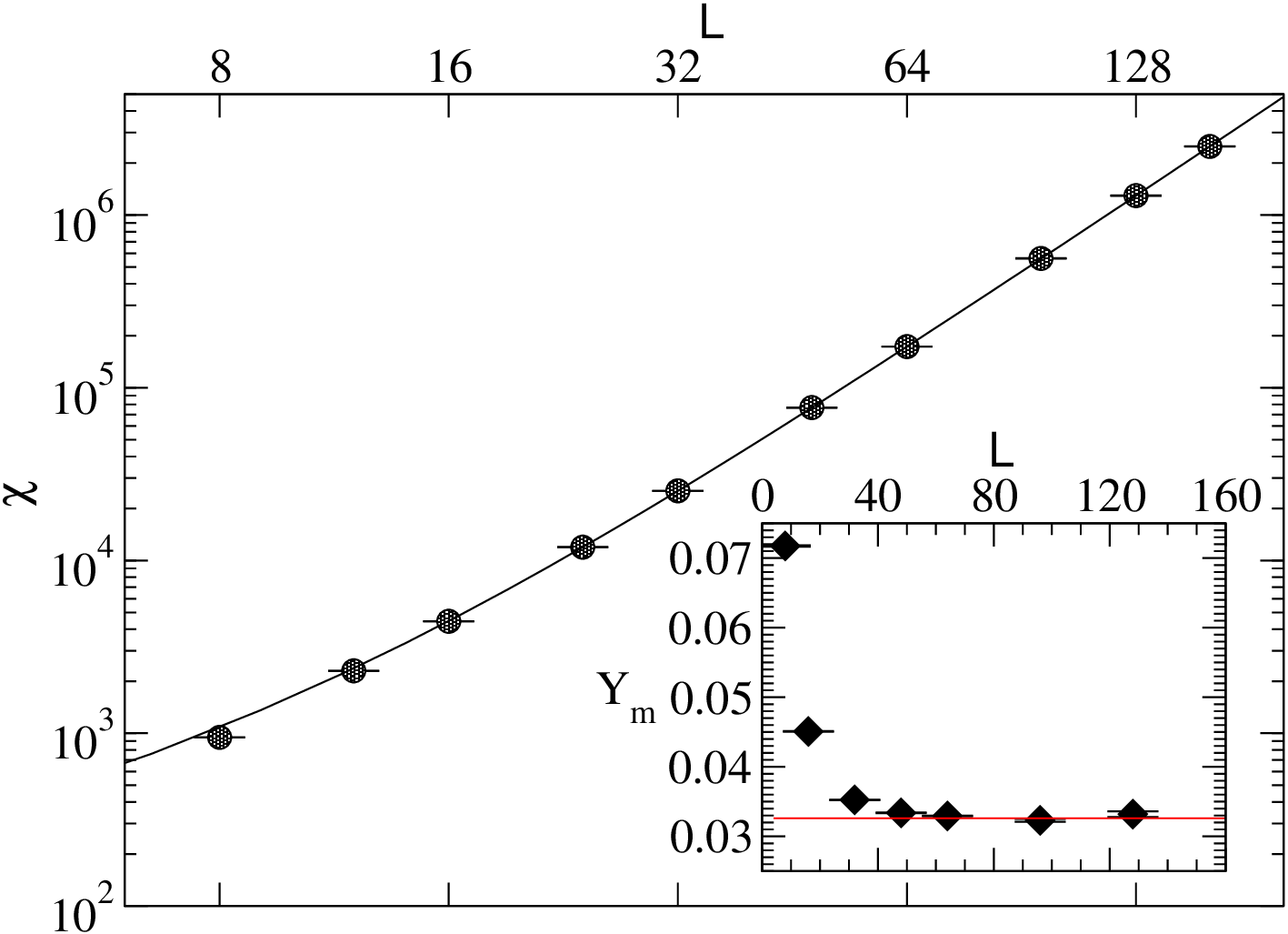}
\caption{Plot of $\chi$ vs. $L$ and $Y_m$ vs. $L$ (inset) 
at $T=7.42$ and $m=0$.}
\label{fig1}
\end{figure}

We have done extensive calculations on various lattice sizes in 
the range $8 \leq L \leq 192$ with $L_t=4$. In order to understand
the critical behavior we have measured $\chi$ for $m=0$ at 
different values of $T$ between $7.3$ and $7.5$. On the other hand 
we focused on a single temperature in the broken phase ($T=7.42$)
and computed $\cc,Y_m$ and $M_\pi$ for masses in the range 
$0 \leq m \leq 0.01$. In figure \ref{fig1} we plot our results for
$\chi$ as a function of $L$ at $m=0$ and $T=7.42$. The data fits well to 
the ChPT prediction  (eq.(\ref{chptchi})) with $\Sigma = 1.079(2)$, 
$F = 0.181(4)$ and $a=114(4)$ with a $\chi^2$/d.o.f of 0.73.
The value of $F$ obtained from this fit is in excellent agreement with
$F=0.181(1)$ obtained through a direct evaluation of $Y_m$ at $m=0$ at 
large volumes. This can be seen from the plot shown in the inset of
figure \ref{fig1}.
\begin{figure}[htb]
\vspace{9pt}
\includegraphics[width=17pc]{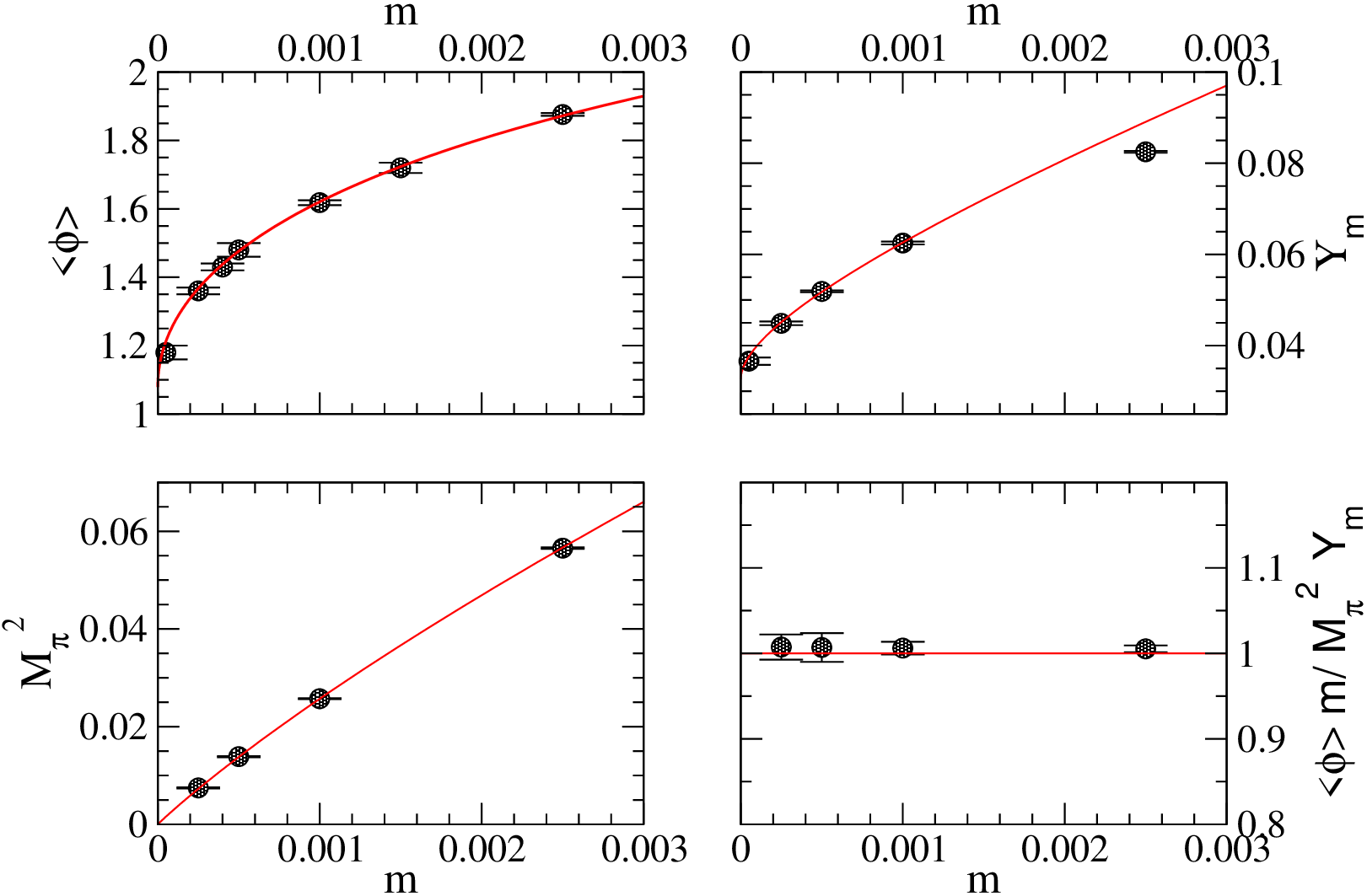}
\caption{Plots of $\cc$ (upper-left) $Y_m$ (upper-right), $M_\pi$
(lower-left), extrapolated to $L=\infty$, as a function of the quark 
mass. Check of the Gellmann-Oaks-Renner relation (bottom-right).}
\label{fig2}
\end{figure}

Given this excellent agreement with ChPT at $m=0$ we have also looked for
the quark mass dependence of $\cc$, $Y_m$ and $M_\pi$. These quantities 
were computed for various volumes until their values did not change for 
two different volumes. This ``infinite'' volume data is plotted on figure
\ref{fig2}.
When the data is fit to the prediction from ChPT 
(eqs.~(\ref{ccm}),(\ref{ym}) and (\ref{chptmass})), 
with $\Sigma$ and $F$ fixed to $1.079$ and $0.181$ (see above), we find 
that $\alpha_1=17.9(2)$, $\alpha_2 = -64(3)$, $\alpha_3=20(1)$, 
$\alpha_4=301(36)$, $\alpha_5 = -8.14(10)$ and $\alpha_6 = 38(2)$.
The $\chi^2$/d.o.f of all the fits are close to one. Although these
fits appear to be good, the results are not consistent with ChPT.
In particular ChPT predicts $\alpha_3 = 0$ \cite{Has90}. In spite 
of this disagreement, the bottom right plot of figure \ref{fig2} shows 
that our results satisfy the the Gellmann-Oaks-Renner relation 
(eq.((\ref{gorr})). 

The only plausible explanation we can imagine for the disagreement, is that 
most of the masses used in the fit are perhaps too heavy for ChPT
to be valid. This is because the cutoff in ChPT is proportional to $F^2$ 
(we are effectively in three dimensions), which is quite small in our case. 
Indeed the coefficients $\alpha_i$ are also uncomfortably large for the 
same reason. We are currently studying another temperatures deeper inside 
the broken phase where $F^2$ is larger.

\begin{figure}[htb]
\vspace{9pt}
\includegraphics[width=17pc]{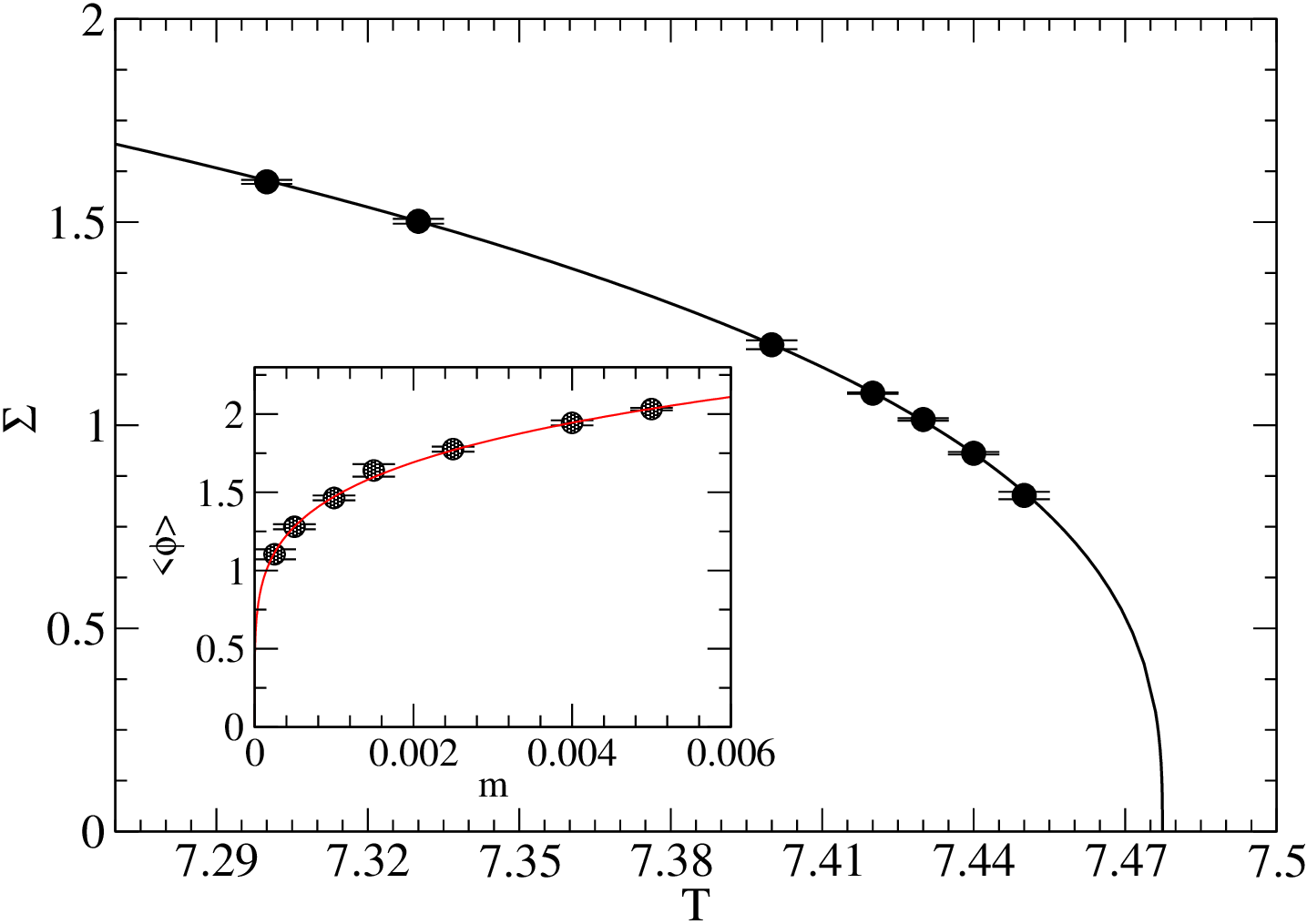}
\caption{ Plots of $\Sigma$ vs. $T$ and $\cc$ vs. $m$ at $T=T_c$
(inset).}
\label{fig3}
\end{figure}

Finally a careful analysis of the finite size effects of $\chi$
at various temperatures reveals that $T_c=7.47739(3)$. We also find
that the data for $\Sigma$ and $\cc$ fits well to the form
predicted by eqs.~(\ref{sigma}) and (\ref{crit}). We find 
$\beta=0.348(2)$, $A=2.92(2)$ with a $\chi^2/$d.o.f of 0.53 for the 
fit to eq.~(\ref{sigma}) and $\delta=4.97(10)$, $B=5.9(1)$ with a 
$\chi^2/$d.o.f of 0.34 for the fit to eq.~(\ref{crit}). The results 
and the fits are shown in figure \ref{fig3}.

\end{document}